\documentclass[preprint,showpacs,preprintnumbers,amsmath,amssymb]{revtex4}

\usepackage{graphicx}
\usepackage{dcolumn}
\usepackage{bm}
\usepackage{amsmath,dsfont}
\usepackage{amssymb,amsthm,amscd, amsbsy, array}
\usepackage{graphics,graphicx,xcolor}
\usepackage{etex}
\usepackage[all]{xy}
\usepackage{hyperref}

\numberwithin{equation}{section}

\newcommand{\bflr}{\begin{flushright}}
\newcommand{\eflr}{\end{flushright}}
\newcommand{\bc}{\begin{center}}
\newcommand{\ec}{\end{center}}
\newcommand{\ben}{\begin{enumerate}}
\newcommand{\een}{\end{enumerate}}

\newcommand{\be}{\begin{equation}}
\newcommand{\ee}{\end{equation}}
\newcommand{\ba}{\begin{array}}
\newcommand{\ea}{\end{array}}

\newcommand{\dd}[2]{\frac{\partial{#1}}{\partial{#2}}}

\newcommand{\rg}{\rho}

\newcommand{\vf}{\varphi}

\newcommand{\Gam}{\Gamma}

\newcommand{\ld}{\left.}

\newcommand{\der}{\partial}

\newcommand{\half}{{\scriptstyle{\frac{1}{2}}}}
\def\2{{\half}}
\def\smallcirc{{\,\raise 0.5pt \hbox{$\scriptstyle\circ$}\,}}
\newcommand{\const}{\mathop{\rm const}\nolimits}

\def\p{{\partial}}

\def\beq{\begin{equation}}
\def\eeq{\end{equation}}
\def\beqa{\begin{eqnarray}}
\def\eeqa{\end{eqnarray}}
\def\nn{\nonumber}
\def\barray{\left(\begin{array}}
\def\earray{\end{array}\right)}
\def\barraynb{\begin{array}}
\def\earraynb{\end{array}}
\def\benu{\begin{enumerate}}
\def\eenu{\end{enumerate}}
\def\bea{\begin{eqnarray}}
\def\eea{\end{eqnarray}}

\def\smallover#1/#2{\hbox{$\textstyle\frac{#1}{#2}$}} %

\def\vY{{\bm{Y}}}

\def\vnabla{{\overrightarrow{\nabla}}}

\newcommand{\sgn}{\mathop{\mathrm{sgn}}}

\def\benu{\begin{enumerate}}
\def\eenu{\end{enumerate}}

\usepackage{color}

\newcommand{\dm}{d} 
\newcommand{\h}[1]{\hat{#1}} 
\newcommand{\F}{\mathcal{F}} 
\newcommand{\wh}[1]{\widehat{#1}}

\begin{document}

\title{Conformal Killing Tensors and covariant Hamiltonian Dynamics}

\author{
M. Cariglia$^{1}$\footnote
{e-mail:marco@iceb.ufop.br}
G.~W.Gibbons$^{2,3,4}$\footnote{mail:G.W.Gibbons@damtp.cam.ac.uk},
J.-W. van Holten$^{5,6}$\footnote{
mail: t32@nikhef.nl},
P.~A.~Horvathy$^{4,7}$\footnote{
e-mail:horvathy-at-lmpt.univ-tours.fr},
P.-M. Zhang$^{4,7,8}$\footnote{e-mail:zhpm@impcas.ac.cn}
}

\affiliation{
$^{1}$DEFIS, Universidade Federal de Ouro Preto, Campus Moro de Cruzeiro, 35400-000 Ouro Preto, MG-Brasil
\\
$^{2}$Department of Applied Mathematics and Theoretical  Physics,
Cambridge University, Cambridge, UK
\\
$^{3}$LE STUDIUM, Loire Valley Institute for Advanced Studies, Tours and Orleans France
\\
$^{4}${\it Laboratoire de Math\'ematiques et de Physique Th\'eorique}, Tours University
(France)
\\
$^{5}$NIKHEF, Amsterdam (Netherlands)
\\
$^{6}$ Leiden University, Leiden (Netherlands)
\\
$^{7}$
Institute of Modern Physics, Chinese Academy of Sciences,
Lanzhou (China)
\\
$^{8}$
State Key Laboratory of Theoretical Physics, Institute of Theoretical Physics, Chinese Academy of Sciences, Beijing 100190, China 
}
\date{\today}

\begin{abstract} 
A covariant algorithm for deriving the conserved quantities for natural Hamiltonian systems is 
combined with the non-relativistic  
 framework of  Eisenhart, and of Duval, in which the classical  trajectories arise
as geodesics in a higher dimensional space-time,
 realized by Brinkmann manifolds.
Conserved quantities which are polynomial in the momenta can be built using 
time-dependent conformal Killing tensors with flux. The latter are associated with terms proportional to the Hamiltonian in the lower dimensional theory and with spectrum generating algebras for higher dimensional quantities of order $1$ and $2$ in the momenta. 
 Illustrations of the general theory include the Runge-Lenz vector for planetary motion with a time-dependent gravitational constant $G(t)$, 
 motion in a time-dependent electromagnetic field of a certain form, quantum dots, the H\'enon-Heiles and 
  Holt systems, respectively, providing us with Killing tensors of rank that ranges from one to six.  \\
\bigskip
\noindent
KEY WORDS:\\ Time-dependent symmetries, conserved quantities, Conformal Killing tensors,
 Covariant Dynamics. 
\end{abstract}

\pacs{
\\
45.05.+x,  General theory of classical mechanics of discrete systems 
\\
11.30.-j 	Symmetry and conservation laws
\\
11.30.Na, Nonlinear and dynamical symmetries (spectrum-generating symmetries) 
\\ 
}

\preprint{arXiv:1404.342v4 [math-ph]
} 
\maketitle

\tableofcontents

\section{Introduction}
Properties like separability and integrability of  Hamiltonian systems which are quadratic in the momenta (referred to as ``natural ones'' \cite{Benenti2001}), can conveniently be
studied in terms of Killing tensors \cite{KalninsMiller1980,KalninsMiller1981,Benenti1997,Benenti2001,Benenti2005}; the latter are also used to  explain hidden symmetries \cite{Carter,Crampin,DuvalValent}.  
The most common application of Killing tensors is to Runge-Lenz-type
quantities which are themselves quadratic in the momenta, but
 higher rank Killing tensors with flux were also considered  \cite{GaryDavidClaudeHouri2011,Galajinsky,Visinescu,Marco2012,Rugina,CGvHHKZ2014}. 

Conserved quantities which are polynomial in the momenta can be searched for systematically  using the covariant approach in \cite{gibbons-rietdijk-vholten1993,jwvholten2006},
later extended to the non-Abelian context \cite{PeterNgome2009}, to curved spaces \cite{Ngome,Visinescu2009,Igata}, to spinning particles \cite{Kubiznak} and to supersymmetry \cite{PeterJWNgome2010}.

Natural Hamiltonian systems can be lifted to a \emph{higher dimensional Kaluza-Klein space-time endowed with a Lorentz metric and a covariantly constant null vector}, 
originally introduced by Eisenhart \cite{Eisenhart}.  
 Such manifolds had in fact already been studied   by Brinkmann \cite{Brinkmann1925} in a general relativistic context, but with a rather different aim: Brinkmann was interested to find all Einstein manifolds which are conformal to Einstein manifolds. Eisenhart's framework was 
 rediscovered, independently, in 1984 by Duval et al. \cite{DBKP,DGH91}
in their search for a geometric framework for understanding the action of the one-parameter central extension of the Galilei group called the Bargmann group \cite{SSD}. Therefore they called it a ``Bargmann space''.

In this framework, the non-relativistic motions are  obtained by  projecting the null geodesics of Bargmann space into non-relativistic space-time.
 
 The geodesic formulation, besides being simple and elegant, is particularly useful for integrable systems as there exists a well established theory of the separation of variables for the Hamilton-Jacobi and the Schr\"{o}dinger equation for geodesic Hamiltonians \cite{KalninsMiller1980,KalninsMiller1981,Benenti1997,Benenti2005}. 
It is also convenient to study the Newton-Cartan theory of gravity \cite{DBKP}, and can include components interpreted as vector potentials \cite{DHP2}. 
 
The Bargmann framework has many applications: a non-exhaustive list includes  the Schr\"odinger symmetry \cite{Jackiw1972,Niederer1972,Hagen1972,Duval0,DBKP},
the study of Dirac type gravitational theories with a time dependent gravitational constant $G(t)$ as suggested by Dirac \cite{Dirac} and of the two types of non-relativistic theories of electromagnetism  \cite{DGH91}, Chern-Simons vortices \cite{DHP2,DHP1}, applications to hydrodynamics \cite{HHhydro}, Goryachev-Chaplygin-Kovalevskaya tops
\cite{Rugina},
the non-relativistic AdS/CFT correspondence \cite{DuvalLazzarini2012}, the Dirac equation with flux \cite{Marco2012}, Toda chains \cite{GaryMarco2013} etc.
For time-independent systems a second type of higher dimensional description is also possible,
as noticed already by Eisenhart \cite{Eisenhart}.  
 
In this paper we combine these two, complementary techniques by lifting the covariant framework to Bargmann space, upon which the previously considered Killing tensors are promoted to conformal ones. The lifts of conserved quantities that are polynomial in the momenta for the original system correspond to conserved higher dimensional quantities that are homogeneous in the momenta and are built using conformal Killing tensors.
This provides us with a general theory which allows for discussing conserved quantities associated to time-dependent and/or higher-order symmetries. In particular, a conserved quantity $C$ in phase space that is explicitly time dependent satisfies the equation 
\be 
\frac{dC}{dt} = \frac{\partial C}{\partial t} + \{ C , H \} = 0 \, , 
\ee 
with $\frac{\partial C}{\partial t} \neq 0$, and this in turn implies that it generates a canonical transformation that does not preserve the energy. Such transformations are related to spectrum generating algebras \cite{Dothan}.
 
The main contribuition of this work is discussing in detail the explict time dependence of conserved quantities and tensors: in particular we are able to show that higher dimensional Killing tensors that are properly conformal are related to lower dimensional quantities with terms proportional to the Hamiltonian, and to explicitly time dependent conserved quantities. We check this explicitly for lifted conserved quantities of order $1$ and $2$ in the momenta, and argue on general grounds that the same happens for higher order quantities. 

Standard Killing tensors  are particular cases of conformal ones: while  a conformal Killing tensor generates a canonical transformation that multiplies the higher dimensional Hamiltonian by a non-trivial factor,  a Killing tensor leaves the Hamiltonian invariant, see (\ref{2.22})  and below. 

We then illustrate our general theory with various examples: we give examples of conformal Killing tensors in the case of a free particle, of a Dirac-type of theory of gravity with time dependent gravitational constant and in the case of a theory with time dependent electric and magnetic fields that generalises earlier considerations of Lynden-Bell. We also consider cases with standard Killing tensors, such as the quantum dot, the H\'enon-Heiles and the Holt system.

\section{Eisenhart-Duval lift and the covariant formalism\label{sec:geodesic_lift}} 
 
\subsection{Bargmann space and Brinkmann metric\label{sec:Brinkmann}}

The non-relativistic Kaluza-Klein-type  spacetime of Eisenhart \cite{Eisenhart} describes non-relativistic physics 
 by an $n=d+2$ dimensional manifold endowed with a Lorentz metric and a covariantly constant Killing vector. Let us  describe  briefly such metrics, independently of whether they solve Einstein's equations or not.
The $n=d+2$-dimensional Brinkmann line element is 
\be \label{eq:Brinkmann_metric} 
ds^2 =g_{\mu\nu} dx^\mu dx^\nu = h_{ij} dx^i dx^j + 
 2 du \left( dv - \Phi du + N_i dx^i \right) \, , \quad i, j = 1, \dots, d  \, . 
\ee 
The components $(\Phi, N_i; h_{ij})$ are independent of the coordinate $v$: 
\be 
\partial_v \Phi = \partial_v N_i = \partial_v h_{ij} = 0 \, . 
\ee 
In what follows, we work in a $n = d + 2$ split formulation; space indices are raised and lowered by the transverse metric $h_{ij}$ and its inverse $h^{ij}$.

The metric is invariant under local transformations 
\be \label{eq:Brinkmann_invariance}
\delta v = \Lambda(u, x^i) \, , \quad \delta \Phi = \partial_u \Lambda (u, x^i) \, , 
\quad \delta N_i = - \partial_i \Lambda (u, x^i) \, , \quad \delta h_{ij} = 0 \, . 
\ee 
The geodesic Lagrangian is
\be 
L = \frac{1}{2} g_{\mu\nu} \dot{x}^\mu \dot{x}^\nu = \frac{1}{2} h_{ij} \dot{x}^i \dot{x}^j + \dot{u} \dot{v} - \Phi \, \dot{u}^2 + N_i \, \dot{x}^i \, \dot{u} \, , 
\ee 
where $\dot{x}^\mu = {dx^\mu}/{d \lambda}$ with $\lambda$ an affine geodesic parameter.
The geodesic hamiltonian which parameterises the inverse metric is 
\bea 
\mathcal{H} &=& \frac{1}{2} g^{\mu\nu} \h{p}_\mu \h{p}_\nu =  \h{p}_u \h{p}_v + \left(\Phi + \frac{1}{2} h^{ij} N_i N_j \right) \h{p}_v^2 - h^{ij} N_j \h{p}_i \, \h{p}_v + \frac{1}{2} h^{ij} \h{p}_i \h{p}_j \nn \\ 
&=&  \h{p}_u \h{p}_v + \Phi \h{p}_v^2 + \frac{1}{2} h^{ij} \left(\h{p}_i - N_i \, \h{p}_v \right) \left(\h{p}_j - N_j \, \h{p}_v \right) \, . 
\eea 
The conjugate momentum, $\h{p}_v$ of the  a cyclic coordinate  $v$  is  conserved along geodesics, 
$ 
{d \h{p}_v}/{d \lambda} = - {\partial \mathcal{H}}/{\partial v} = 0 \, , 
$ 
where $\lambda$ is the affine geodesic parameter. Then we can write 
\be 
\h{p}_v =  q = \const , \; \mathcal{H} = q  \h{p}_u + H, 
\qquad
H = \frac{1}{2} h^{ij} \left(\h{p}_i - q N_i  \right) \left(\h{p}_j - q N_j \right) + q^2 \Phi \, . 
\label{eq:H} 
\ee  
 This is the most general example of natural Hamiltonian in $\dm$ dimensions, where $\Phi$ and $N_i$ play the role of a scalar and, respectively, a vector potential. When the potentials take an appropriate form, the Hamiltonian $H$ can be invariant under supersymmetry transformations: these are related to symmetries of the theory of the spinning particle moving in the $h$, $N$, $\Phi$ background, and to the presence of (conformal) Killing-Yano tensors \cite{gibbons-rietdijk-vholten1993}. A discussion of Killing-Yano tensors for the Eisenhart lift geometry can be found in \cite{Marco2012}. 

For $q \neq 0$ the affine parameter $\lambda$ can be eliminated by observing that 
$
{du}/{d\lambda} = {\partial \mathcal{H}}/{\partial \h{p}_u} =  \h{p}_v = q \; \Rightarrow \; d\lambda = q^{-1} du \, . 
$
We will consider only light-like geodesics, for which 
\be \label{eq:null_condition}
\mathcal{H} = 0, \quad\hbox{i.e.} 
\quad
q \h{p}_u = - H \, . 
\ee 
Then the equations of motion are given by 
\bea 
\frac{dv}{d\lambda} &=&  \frac{\partial \mathcal{H}}{\partial \h{p}_v} = - \frac{1}{q} \left[ \frac{1}{2} h^{ij} \left(\h{p}_i - N_i \, \h{p}_v \right) \left(\h{p}_j - N_j \, \h{p}_v \right) - q^2 \Phi \right] - h^{ij} \left( \h{p}_i - N_i \, \h{p}_v \right) N_j \, , 
\label{eq:v_eom}
\\[4pt] 
\frac{d \h{p}_u}{d\lambda} &=&  - \frac{\partial \mathcal{H}}{\partial u} \, ,  
\label{eq:pu_eom} 
\\[4pt] 
\frac{d x^i}{d\lambda}  &=& \frac{\partial \mathcal{H}}{\partial \h{p}_i}   = h^{ij} (\h{p}_j -  N_j \, \h{p}_v)  \, ,  
\label{eq:xi_eom} 
\\[4pt] 
\frac{d \h{p}_i}{d\lambda}  &=& -  \frac{\partial \mathcal{H}}{\partial x^i} \, . 
\label{eq:pi_eom}
\eea 
 Eq. \eqref{eq:pu_eom} is solved by \eqref{eq:null_condition}, while eqs.\eqref{eq:xi_eom}, \eqref{eq:pi_eom} become the equations of motion for a $\dm$-dimensional system with Hamiltonian \eqref{eq:H} upon fixing $\h{p}_v = q$. Therefore, the full equations of motion in $n = \dm + 2$ dimensions restricted to such geodesics are equivalent to the equations of motion for the $\dm$ 
 dimensional system
 with coordinates $x^i$ and $u$ viewed as time, augmented with \eqref{eq:null_condition} and
  \eqref{eq:v_eom}. The latter
 can be integrated once the functions $x^i(\lambda)$, $\h{p}_j(\lambda)$ have been determined. In fact, identifying $H$ in eqn.\eqref{eq:null_condition}  with the generator of time translations for the $\dm$-dimensional system yields
\be 
t = - \frac{1}{q} u = - \lambda \, . 
\ee 
These relations are important since  we  want to project the $\dm +2$-dimensional system on a $d+1$-dimensional base where the natural coordinates $(x^i, t)$ will be used. In particular, we will need to distinguish between the higher dimensional momenta $\hat{p}_i = {\partial L}/{\partial \dot{x}^i} = h_{ij} \dot{x}^i + \dot{u} N_i = h_{ij} \dot{x}^i + q N_i$, and a lower dimensional version defined using the time variable: $p_i = {\partial L}/{\partial ({d x^i}/{dt})} = -\hat{p}_i =   h_{ij} {d x^i}/{dt} - q N_i$. In terms of the latter, 
 \be \label{eq:H_lower_d}
H = \frac{1}{2} h^{ij} \left(p_i + q N_i  \right) \left(p_j + q N_j \right) + q^2 \Phi \, . 
\ee 
Under a gauge transformation $\delta N_i = \partial_i \mu (u, x)$, $\delta \Phi = - \partial_u \mu (u, x)$, the right hand side of \eqref{eq:v_eom} changes by $ - {d \mu}/{d\lambda}$, which can be re-absorbed by the transformation $\delta v = - \mu$, included into  
\eqref{eq:Brinkmann_invariance}. Therefore we can build a $U(1)$ vector potential as follows. Allowing $\Phi$ to be of the form $\Phi_1 + \Phi_2$, where under a gauge transformation $\delta \Phi_2 = 0$, we can set 
\be 
A = N_i dx^i - \Phi_1 du \, , 
\ee 
so that under a gauge transformation $\delta A = d \mu$. The $U(1)$ field strenght is given by 
\bea  
\mathcal{F} = dA &=& \frac{1}{2} \left( \partial_i N_j - \partial_j N_i \right) dx^i \wedge dx^j + \left( \partial_u N_i + \partial_i \Phi_1 \right) du \wedge dx^i \nn \\ 
&=& \frac{1}{2} \left( \partial_i N_j - \partial_j N_i \right) dx^i \wedge dx^j + \left( \partial_t N_i - q \partial_i \Phi_1 \right) dt \wedge dx^i 
\, . 
\eea 
With this one can proceed as in \cite{DGH91} and obtain the two types of Galilean electromagnetic theories in $\dm+1$ dimensions.
  
It is worth noticing that the right hand side of \eqref{eq:v_eom} changes under a gauge transformation by $ - {d \mu}/{d\lambda}$, which can be re-absorbed by the transformation $\delta v = - \mu$, included into  
\eqref{eq:Brinkmann_invariance}. 

Let the manifold $\wh{\mathcal{P}}$ be  spanned by the variables $(x^\mu,\h{p}_\nu)$, typically the cotangent bundle of a base manifold $\wh{\mathcal{M}}$. The covariant derivatives are defined by
\be \label{eq:covariant_derivative}
\wh{D}_\mu f \equiv \ld \dd{f}{x^\mu} \right|_p + \wh{\Gam}_{\mu\nu}^{\,\rho}\, \h{p}_\rho\, \dd{f}{\h{p}_\nu} \, , 
\ee 
where $\wh{\Gam}$ are the Christoffel symbols of the metric $g$. Due to the symmetry of the latter, the Poisson brackets  can be written covariantly as \cite{gibbons-rietdijk-vholten1993,jwvholten2006,Visinescu2009}
\be \label{eq:Poisson_covariant} 
\left\{f, g \right\}_{\wh{\mathcal{P}}} =
 \p_\mu f \dd{g}{\h{p}_\mu} - \dd{f}{\h{p}_\mu} \p_\mu g=
  \wh{D}_\mu f \dd{g}{\h{p}_\mu} - \dd{f}{\h{p}_\mu} \wh{D}_\mu g\,.
\ee 

To clarify the action of $\wh{D}_\mu$, any phase space function that admits a power expansion in the momenta with non-negative exponents will contain terms of the kind $T(q)^{\nu_1 \dots \nu_p} \h{p}_{\nu_1} \dots \h{p}_{\nu_p}$, where $T$ is a tensor. $\wh{D}_i$ acts on such terms as $\wh{\nabla}_\mu T^{\nu_1 \dots \nu_p} \h{p}_{\nu_1} \dots \h{p}_{\nu_p}$, where $\wh{\nabla}$ is the Levi-Civita covariant derivative acting on tensors. Similar equations for covariant derivatives of phase space functions have been used in \cite{DavidPavelValeriMarco2012} in order to give a geometrical description of Lax pairs associated to phase space tensors that are constant along trajectories. 

The definitions above reproduce  the canonical equations of motion. First of all, 
\be 
\frac{dx^\mu}{d\lambda} = \left\{ x^\mu, \mathcal{H} \right\}_{\h{\mathcal{P}}} = \dd{\mathcal{H}}{\h{p}_\mu}  \, , 
\ee 
which implies $\h{p}_\mu = g_{\mu\nu}{dx^\nu}/{d\lambda}$ . 
Next, writing for a generic phase space function $f(x,\h{p})$ 
\be 
\big\{f, \mathcal{H} \big\}_{\h{\mathcal{P}}} = \frac{d f}{d\lambda} = \frac{\partial f}{\partial x^\mu} \frac{d x^\mu}{d\lambda} + \frac{\partial f}{\partial \h{p}_\mu} \frac{d \h{p}_\mu}{d\lambda} \, , 
\ee 
and equating this with the covariant expression that comes from \eqref{eq:Poisson_covariant}, one gets 
\be 
\frac{\h{D}\h{p}_\mu}{D\lambda} \equiv \frac{d\h{p}_\mu}{d\lambda} - \frac{dx^\nu}{d\lambda}\, \wh{\Gam}_{\mu\nu}^{\, \rho}\, \h{p}_\rho = 0 \, , 
\ee 
which is the \emph{geodesic equation}. 
 
In the framework we propose here, conserved quantities   that are polynomial in the momenta are built from \emph{conformal Killing tensors}. Let us explain how. We assume that a conserved quantity can be written as 
\be \label{eq:power_expansion}
\wh{C} = \sum_{r \geq 0} \h{C}_{(r)} = \sum_{r \geq 0} \frac{1}{r!} \wh{C}_{(r)}^{\, \mu_1 .... \mu_r}(x) \h{p}_{\mu_1} ... \h{p}_{\mu_r} \, . 
\ee 
Requiring that the quantity is conserved for null geodesics leads to a set of decoupled equations
\be
\label{2.22} 
\left\{\wh{C}_{(r)}, \mathcal{H}\right\}_{\h{\mathcal{P}}} = f(x,\h{p})_{(r-1)} \mathcal{H} \, , 
\ee 
where $f(x, \h{p})_{(r-1)} = \frac{1}{(r-1)!} f(x, \h{p})_{(r-1)}^{(\lambda_1 \dots \lambda_{r-1})}  \h{p}_{\lambda_1} \dots \h{p}_{\lambda_{r-1}}$. In tensorial notation this is equivalent to 
\be \label{eq:conformal_Killing} 
\wh{C}_{(r)}^{(\mu_1 ..i_r; \mu_{r+1})} = \frac{r}{2} f_{(r-1)}^{(\mu_1 \dots \mu_{r-1}} g^{\mu_r \mu_{r+1})} \, ,  
\ee
and therefore $f_{(r-1)}$ is related to the divergence of $\wh{C}_{(r)}$ and to derivatives of its trace by taking the trace of \eqref{eq:conformal_Killing}. This is the definition of a \textit{conformal Killing tensor}. In the special case when $f=0$ then $\h{C}_{(r)}$ reduces to a usual Killing tensor: the canonical transformation associated to it leaves the Hamiltonian $\cal{H}$ invariant.

Further details will be presented in the next subsections. 

\subsection{The Natural Hamiltonian}\label{NatHam}

The Hamiltonian \eqref{eq:H} can be rewritten in terms of covariant momenta $\Pi_i = p_i + q N_i$ as 
\be \label{eq:natural_Haimiltonian} 
H = \frac{1}{2}\, g^{ij}\, \Pi_i \Pi_j + q^2 \Phi  \, .  
\ee 

We now relate the $n$-dimensional and $\dm$ dimensional quantities. Let $\mathcal{P}$ be the manifold spanned by $(x^i, p_j)$. 
The covariant geodesic Poisson brackets \eqref{eq:Poisson_covariant} can be re-written using the $(x^i, v, u)$ split. Knowing the form of the non-zero Christoffel symbols of $g_{\mu\nu}$ (given in the Appendix) and using the identity 
\be 
\left. \dd{f}{x^i} \right|_{p}= \left. \dd{f}{x^i} \right|_{\Pi} + q \frac{\partial f}{\partial \Pi_j} \partial_i N_j \, , 
\ee  
we find that for any two functions $f (q^i, v, u, p_i, p_v, p_t)$, $g (q^i, v, u, p_i, p_v, p_t)$,
\be 
 \left\{f, g \right\}_{\h{\mathcal{P}}} = -\left\{f, g\right\}_{\mathcal{P}} + \frac{\partial f}{\partial v} \frac{\partial g}{\partial p _v} - \frac{\partial f}{\partial p_v} \frac{\partial g}{\partial v} + 
\frac{\partial f}{\partial u} \frac{\partial g}{\partial p_u} - \frac{\partial f}{\partial p_u} \frac{\partial g}{\partial u} \, , 
\ee  
where the covariant Poisson brackets on $\mathcal{P}$ are given by 
\be \label{eq:Poisson_covariant_downstairs} 
\left\{f, g \right\}_{\mathcal{P}} = D_i f \dd{g}{\Pi_i} - \dd{f}{\Pi_i} D_i g  - q \F_{ij}  \dd{f}{\Pi_i} \dd{g}{\Pi_j} \, .  
\ee
Here $\F_{ij} = \der_i N_j - \der_j N_i$ is the field-strength tensor of the  vector potential $N_i$  and the covariant derivatives are defined by
\be
D_i f \equiv \ld \dd{f}{q^i} \right|_{\Pi} + \Gam_{ij}^{\,k}\, \Pi_k\, \dd{f}{\Pi_j} \, , 
\ee  
where the $\Gam$s are the Christoffel coefficients of the metric $h$. 
Thus by  studying the higher dimensional geodesics we obtain the known covariant form of dynamics for natural Hamiltonians that has been already discussed in the literature \cite{gibbons-rietdijk-vholten1993,jwvholten2006,Visinescu2009,PeterNgome2009,PeterJWNgome2010}. 

For natural Hamiltonians a function $C$ of the form 
\be 
C = \sum_{r \geq 0} \frac{1}{r!} C_{(r)}^{\, i_1 .... i_r}(x) \Pi_{i_1} ... \Pi_{i_r} \,  
\ee 
is  conserved  if and only if the tensors $C_{(r)}$ satisfy a set of coupled Killing equations with flux \cite{gibbons-rietdijk-vholten1993,jwvholten2006}. 

Some time ago, Crampin \cite{Crampin} considered a special case of these equations that holds when $C$ is written as a scalar term plus a term of order $q \ge 1$ in the momenta, in particular he wrote the Runge-Lenz vector in the Kepler problem.  
 
In \cite{GaryDavidClaudeHouri2011} it has been shown that a polynomial conserved quantity of the form 
\be 
C = \sum_{r = 1}^m C_{(r)} \,  
\ee 
where each $C_{(r)} =C_{(r)}(u, x^i,p_j) $ is a homogeneous polynomial of degree $r$ in the momenta, lifts to a conserved quantity for the lifted Hamiltonian $\mathcal{H}$ that is homogeneous of degree $m$ and is written as 
\be \label{eq:conserved_quantity_lift} 
\wh{C} = \sum_{r = 1}^m \left( \frac{\h{p}_v}{q}\right)^{m-r} C_{(r)} \, , 
\ee 
therefore being associated to a conformal Killing tensor. Asking for $\{ C, H \}_{\mathcal{P}} = 0$ splits into a set of conditions for each order $a$ in the momenta  
\be \label{eq:conservation_order_a}
\frac{\partial C_{(r)}}{\partial u} + \big\{ C_{(r-1)}, H_{(2)} \big\} + \big\{ C_{(r)}, H_{(1)} \big\} + \big\{ C_{(r+1)}, H_{(0)} \big\} = 0 \, , 
\ee 
where $H_{(0,1,2)}$ are the homogeneous in the momenta parts of $H$ that are of order $0$, $1$ and, respectively, $2$. 
 
In terms of the $\Pi$ variables, if we write $C = C(x^i, \Pi_j ) = \sum_{r = 1}^m C^{\prime}_{(r)}(x^i,\Pi_j)$ for new coefficients $C^{\prime}_{(r)}$, then the  recipe above translates into the simple rule 
\be 
\sum_{r = 1}^m C^{\prime}_{(r)}(x^i, \Pi_j) \rightarrow \sum_{r = 1}^m \left(  \frac{\h{p}_v}{q}\right)^{m-r} C^{\prime}_{(r)} (x^i, -\h{\Pi}_j) \, , 
\ee 
where $\h{\Pi}_i = \hat{p}_i - \, \hat{p}_v \, N_i$.  Note that the $C_{(r)}$ and $C^{\prime}_{(r)}$ tensors above can contain, in particular, the metric $h$ or tensor products of factors of $h$, which  enter equation \eqref{eq:conservation_order_a}. By separating the Killing tensors with flux of \cite{gibbons-rietdijk-vholten1993,jwvholten2006,Visinescu2009,Igata} into a pure metric part and a non-trivial tensorial remainder, we are going to show that the latter satisfies  generalized conformal Killing equations with flux. 
 
In the next two sections we analyse the conservation equation $\{ C , H \}_{\mathcal{P}} = 0$ for quantities related first to conformal Killing vectors, and then generalize to  higher rank tensors. We allow for an explicit $u$-dependence. We start from eqn. \eqref{eq:power_expansion} in order to obtain the right factors of the metric. 
  
\subsection{Conformal Killing vectors\label{sec:conformal_killing_vectors}}

In this subsection we  obtain a generalization of the Killing vector equations  with flux of  \cite{gibbons-rietdijk-vholten1993,jwvholten2006,Visinescu2009}  We will use the following algorithm: 1) write the general form of a conserved quantity in $n = \dm +2$ dimensions that is linear in momenta; 2) project into lower dimension, in particular allowing for $p_u \rightarrow - H/{q}$; 3) this gives a conserved quantity of order 2 in $\dm$ dimensions that naturally has the metric tensor $h$ in the Killing tensor hierarchy; 4) re-writing the standard Killing equations with flux in $\dm$ dimensions for this quantity will automatically yield generalized conformal Killing vector equations. 

In higher dimension a  conserved quantity of order 1 in the 	momenta can be written as 
\be 
\wh{C} = \wh{C}^\mu \h{p}_\mu = \wh{C}^v \h{p}_v + \wh{C}^u \h{p}_u + 
\wh{C}^i \wh{\Pi}_i \, , 
\ee 
which projects into lower dimension as 
\bea 
C &=& - q^2 \wh{C}^v + \wh{C}^u H + q \, \wh{C}^i \Pi_i \, , 
\eea 
which can also be written as 
\be 
C = q^2 ( - \wh{C}^v + \wh{C}^u\Phi ) + \frac{1}{2} \wh{C}^u h^{ij} \Pi_i \Pi_j + q \, \wh{C}^i \Pi_i = C_{(0)} + C^{\, i}_{(1)} \Pi_i + \frac{1}{2} C^{\, ij}_{(2)} \Pi_i \Pi_j
\, .  
\ee 
This yields the identifications 
\be\left\{\begin{array}{lll}
C_{(2)}^{ij} &=& \wh{C}^u h^{ij} \, , 
\\[4pt] 
C_{(1)}^{i} &=&  q \, \wh{C}^i  \, ,
\\[4pt]  
C_{(0)} &=& q^2 ( - \wh{C}^v + \wh{C}^u \Phi ) \, . 
\end{array}\right.
 \label{eq:vector_identifications} 
\ee 
What is really important here is the term with the metric $h$, since it automatically gives the correct Ansatz for the  Killing tensors with flux. 
 
The Killing equations with flux for $C$ generalise to the time-dependent case as 
\be\left\{\begin{array}{lll} 
\nabla^{(i} C^{ \, jk)}_{(2)} &=& 0 \, ,
 \\[4pt] 
\nabla^{(i} C^{ \, j)}_{(1)} &=& - \frac{1}{2}\partial_t C^{\, ij}_{(2)} -q C^{\, (i |l| }_{(2)} \mathcal{F}^{\, j)} {}_l  \, , 
 \\[4pt] 
\nabla^{i} C_{(0)} &=& -\partial_t C^{\, i}_{(1)} - q C^{\, l}_{(1)} \mathcal{F}^{\, i} {}_l + q^2 C^{\, ij}_{(2)} \partial_j \Phi - q C_{(2)}^{ij} \partial_t N_j\, , 
 \\[4pt] 
0 &=& - \partial_t C_{(0)} + q^2 C^i_{(1)} \partial_i \Phi - q C_{(1)}^{i} \partial_t N_i  \, . 
\end{array}\right.
\ee 
Using the identification for $C_{(2)}$ that we get from \eqref{eq:vector_identifications}  this becomes  
\be\left\{\begin{array}{lll}  
\wh{C}^{u} &=& \wh{C}^{u} (t)\, , 
\\[4pt] 
\nabla^{(i} C^{ \, j)}_{(1)} &=&  - \frac{1}{2} \partial_t \left( \wh{C}^{u} h^{ij} \right)   \, , 
\\[4pt] 
\nabla^{i} C_{(0)} &=& - \partial_t C^{\, i}_{(1)} - q C^{\, l}_{(1)} \mathcal{F}^{\, i} {}_l + q^2 \wh{C}^u  \partial^i \Phi_2  - q \wh{C}^u h^{ij} \mathcal{F}_{tj} \, , 
\\[4pt]
0 &=& -  \partial_t C_{(0)} + q^2 C^i_{(1)} \partial_i \Phi_2  - q C_{(1)}^{i} \mathcal{F}_{ti}  \, .
\label{233} 
\end{array}\right.
\ee 
These equations are generalized conformal Killing equations with flux and explicit time dependence. A direct analysis of the Poisson bracket $\{ \h{C} , \cal{H} \}$ shows that there is only one term proportional to the combination $\h{p}_u \h{p}_v$: this has coefficient $\frac{d \h{C}^u}{dt}$. Then, if $\frac{d \h{C}^u}{dt} \neq 0$ it is guaranteed that it must be $\left\{ \h{C} , \cal{H} \right\} = \frac{d \hat{C}^u}{dt} \cal{H}$, and therefore $\h{C}$ is generated by a conformal Killing vector with a $u$-only dependent conformal factor, that commutes with the ``vertical'' Killing vector $\partial_v$. This case applies for example to the free particle of section \ref{sec:free_particle} and to the time-dependent Lorentz-force of section \ref{sec:Lorentz}. 
 
\subsection{Conformal Killing tensors of rank at least two\label{sec:rank2}} 
 
A generic conserved quantity of order $2$  in the momenta can be written on Bargmann space as 
\bea
\wh{C} = \frac{1}{2}\wh{C}^{\mu\nu} \h{p}_\mu \h{p}_\nu 
= \frac{1}{2} \wh{C}^{uu} \h{p}_u^2 + \frac{1}{2}\wh{C}^{vv} \h{p}_v^2 + \wh{C}^{uv} \h{p}_u \h{p}_v  + \wh{C}^{ui} \h{p}_u \h{\Pi}_i + 
\wh{C}^{vi} \h{p}_v \h{\Pi}_i\, + \frac{1}{2} \wh{C}^{ij} \h{\Pi}_i \h{\Pi}_j \, ,\qquad
\eea 
which projects into lower dimension to 
\bea 
C &=& \frac{1}{2} 
\wh{C}^{uu} H^2 + q 
\wh{C}^{ui} H \Pi_i + q^2 ( 
- \wh{C}^{uv} H + \frac{1}{2} 
\wh{C}^{ij} \Pi_i \Pi_j) - q^3 
\wh{C}^{vi}  \Pi_i + \frac{q^4}{2}
\wh{C}^{vv},\quad 
\eea 
also written as 
\bea \label{eq:rank2_ansatz}
C &=&  \frac{1}{8}  
\wh{C}^{uu} h^{ij} h^{kl} \Pi_i \Pi_j \Pi_k \Pi_l + \frac{q}{2} 
\wh{C}^{uk} h^{ij} \Pi_i \Pi_j \Pi_k + \frac{q^2}{2} \left[ \left( 
- \wh{C}^{uv} + 
\wh{C}^{uu} \Phi \right) h^{ij} + 
\wh{C}^{ij} \right] \Pi_i \Pi_j \nn \\ 
&& + q^3 \left( 
- \wh{C}^{vi} + \Phi 
\wh{C}^{ui} \right) \Pi_i + \frac{q^4}{2} \left( 
\wh{C}^{vv} + 
\wh{C}^{uu} \Phi^2 - 2 
\wh{C}^{uv} \Phi \right) \nn \\ 
&=& \frac{d_{(0)}}{4!} h^{ij} h^{kl} \Pi_i \Pi_j \Pi_k \Pi_l + \frac{q}{3!} \, d_{(1)}^k h^{ij} \Pi_i \Pi_j \Pi_k + \frac{q^2}{2} \left[ d_{(2)} h^{ij} + 
\wh{C}^{ij} \right] \Pi_i \Pi_j  + q^3 d_{(3)}^i \Pi_i + q^4d_{(4)}.\qquad \quad 
\eea 
The generic rank-$a$ Killing equations with flux are,
in the time dependent case, for $a \ge 0$, 
\bea \label{eq:generic_rank_Killing_equations}
\nabla^{(i} C^{\, j_1 \dots j_a)}_{(a)} &=& - \frac{1}{a+1} \partial_t \, C_{(a+1)}^{\, i j_1 \dots j_a} - \frac{q}{a+1} C_{(a+2)}^{\, i j_1 \dots j_a l} \partial_t N_l - q \, C_{(a+1)}^{( j_1 \dots j_a l} \mathcal{F}^{i)} {}_l + \frac{q^2}{a+1} \, C_{(a+2)}^{\, i j_1 \dots j_a l} \, \nabla_l \, \Phi \nn \\[6pt] 
&=& - \frac{1}{a+1} \partial_t \, C_{(a+1)}^{\, i j_1 \dots j_a} - \frac{q}{a+1} C_{(a+2)}^{\, i j_1 \dots j_a l} \left( \mathcal{F}_{tl} - q \nabla_l \Phi_2 \right)  - q \, C_{(a+1)}^{( j_1 \dots j_a l} \mathcal{F}^{i)} {}_l \, ,  
\eea 
together with 
\be 
\partial_t \, C_{(0)}+ q  \, C_{(1)}^{\, i } \left( \mathcal{F}_{ti} - q \nabla_i \Phi_2 \right) = 0 \, . 
\ee

So we get the following set of equations: 
\be\left\{\begin{array}{lll}
 d_{(0)} &=& d_{(0)} (t)\, ,
\\[6pt] 
q \nabla^{(i} d_{(1)}^{j_1} h^{j_2 j_3)} &=&  - \frac{1}{4} \partial_t \left( d_{(0)} h^{(i j_1} h^{j_2 j_3)} \right)   \, ,
\\[8pt]
q^2 \nabla^{(i} \left( d_{(2)} h^{jk)} + 
\wh{C}^{jk)} \right) &=& - \frac{q}{3} \partial_t \left( d_{(1)}^{(i} h^{jk)} \right)-\frac{q^2}{3}  d_{(1)}^l \mathcal{F}^{(i} {}_l h^{jk)} - \frac{q d_{(0)}}{3} h^{(ij} \left(\mathcal{F}_t {}^{k)} - q  \nabla^{k)} \Phi_2 \right) \, , 
\\[8pt] 
q^3 \nabla^{(i} d_{(3)}^{j)} &=& - \frac{q^2}{2} \partial_t \left( d_{(2)} h^{ij} + 
\wh{C}^{ij} \right) -q^3 
\wh{C}^{(i |l|} \mathcal{F}^{j)} {}_l \nn \\ 
&& - \frac{q^2}{6} \left[ 2 d_{(1)}^{(i} \left( \mathcal{F}_t {}^{j)} - q  \nabla^{j)} \Phi_2 \right) + d_{(1)}^{\, l} \left(\mathcal{F}_{t l} - q  \nabla_l \Phi_2 \right) h^{ij} \right] \, , 
\\[8pt]
q^4 \nabla^i d_{(4)} &=& - q^3 \partial_t d_{(3)}^{\, i}-q^4 d_{(3)}^{\, l} \mathcal{F}^i {}_l - q^3  \left[ d_{(2)} \left( \mathcal{F}_t {}^i - q \nabla^i \Phi_2 \right) + 
\wh{C}^{i l} \left( \mathcal{F}_{tl} - q \nabla_l \Phi_2 \right) \right] \, ,
\\[8pt] 
0 &=& -  \partial_t d_{(4)} -  d^i_{(3)} \left( \mathcal{F}_{t i} - q \partial_i \Phi \right) \, .
\end{array} \right. 
\label{eq:full_rank2}
\ee 
These equations are generalized conformal Killing equations with flux and explicit time dependence. It is easy to point out a number of cases where the higher dimensional Killing tensor must be properly conformal, by analysing the  Poisson bracket $\{ \h{C} , \cal{H} \}$. If there is a non-zero coefficient $\h{C}^{uu}$ then there is a unique term of the kind $\frac{1}{2} \frac{d\h{C}^{uu}}{dt} \h{p}_u^2 \h{p}_v$: as in the previous section, if the derivative $\frac{d\h{C}^{uu}}{dt}$ is non-zero then the tensor will be conformal. Suppose now that instead $\h{C}^{uu} = 0$ and $\hat{C}^{ui} = 0$, but $\h{C}^{uv} \neq 0$. Then one finds a unique term in the Poisson bracket given by $\left(\h{p}_v \partial_t \h{C}^{uv} + \h{\Pi}_i  \nabla^i \h{C}^{uv}\right) \h{p}_u \h{p}_v$: if this is non-zero then again one will have a conformal Killing tensor. This case applies to the theory of gravity with time-varying coupling constant of section \ref{sec:Gt}. The intermediate case wih $\h{C}^{uu} = 0$,  $\hat{C}^{ui} \neq 0$, $\h{C}^{uv} \neq 0$ is similar in nature but several more sub-terms appear. 
 
Coming to the case of generic rank we note that,
on Bargmann space, a conserved quantity of order $p$ in the momenta can be written as 
\bea
\wh{C} &=& \frac{1}{r!}
\wh{C}^{\mu_1 \dots \mu_r} \h{p}_{\mu_1} \dots  \h{p}_{\mu_r}  \nn \\ 
&=& \sum_{a=0}^r \sum_{b= 0}^{r-a} \frac{1}{a! \, b! \, (r-a-b)!} \, 
\wh{C}^{u_1 \dots u_a v_1 \dots v_b i_1 \dots i_{r-a-b}} \, \h{p}_u^a \, \h{p}_v^b \, \h{\Pi}_{i_1} \dots \h{\Pi}_{i_{r-a-b}} \, . 
\eea 
Then a calculation analogous to the previous ones 
  shows that this reduces ``downstairs"  to 
\bea \label{eq:conserved_quantity_lower_d_generic_p}
C &=& \sum_{l=0}^{2r} \sum_{m=0}^{\min \{l, 2r - l\} } q^{2r-l} \, \frac{d_{(2r - l)}^{\, i_1 \dots i_m}}{l!}  (h)^{\otimes \frac{l-m}{2}}_{j_1 \dots j_{l-m}}  \Pi_{i_1} \dots \Pi_{i_m} \Pi_{j_1} \dots \Pi_{j_{l-m}} \, , 
\eea 
where $l$ is the total degree of the term and $m$ the number of indices of the tensor $d_{(2r-l)}$ (and we set $h^{\otimes \text{half integer number}} = 0$). 
Inserting the explicit form  \eqref{eq:conserved_quantity_lower_d_generic_p} of the coefficients $C_{(a)}$ into equation \eqref{eq:generic_rank_Killing_equations} one gets the generalized conformal Killing equations with flux and explicit time dependence for a generic rank $p$. In particular the highest term in momenta has coefficient $\frac{d_{(0)}(t)}{(2r)!}$, and when it is not constant it is associated to a properly conformal tensor. 

\subsection{The ``d+1''-type  lift} \label{d+1sec}

As explained in sec. \ref{sec:Brinkmann}, the original $\dm$ dimensional system can be recovered from geodesics of the $\dm +2$-dimensional Brinkmann metric by setting $\h{p}_v = q$, $\mathcal{H} = q \h{p}_u + H$. We can also consider a different type of projection when $\Phi$, $N_i$ and $h_{ij}$ do not depend on the $u$ variable. 
Then the $u$ variable can be  eliminated by  a Marsden-Weinstein reduction as follows. $u$-independence implies that   $\partial_u$ is a Killing vector, which generates 
a family of canonical transformations. 
Then $\partial_u$ is the symplectic gradient of $\h{p}_u$. Since $\h{p}_u$ is conserved, the dynamics can therefore be restricted to the $2n-1$--dimensional surface $\Sigma$ defined by $\h{p}_u = 0$ in the symplectic space. The orbits of $\partial_u$ lie in $\Sigma$ since $\partial_u (\h{p}_u) = 0$. Then the quotient of $\Sigma$ by the action of $\partial_u$ is a $2(d+1)$  dimensional symplectic manifold with symplectic form given by the restriction of the original one. The reduced Hamiltonian,
\be  \label{eq:n+1_Hamiltonian} 
H_{\dm+1} = \frac{1}{2} h^{ij} \left( \h{p}_i - \h{p}_v N_i\right) \left( \h{p}_j - \h{p}_v N_j\right) + \h{p}_v^2 \Phi \, ,
\ee 
is quadratic in the momenta and therefore provides us with  a geodesic Lagrangian which comes from the metric 
\be
\label{eq:n+1_metric}
ds_{\dm+1}^2 = h_{ij}dx^i dx^j + \frac{1}{2\Phi}\left(dv + N_i dx^i \right)^2 \, . 
\ee
 $\h{p}_v$ in (\ref{eq:n+1_Hamiltonian}) is a constant and covariant momenta  appear again. 

Under a gauge transformation $\delta N_i = \partial_i \mu(x)$ one has $\delta \left( \dot{x}^i N_i \right) = {d \mu}/{d\lambda}$, which is re-absorbed by $\delta v = - \mu$. 

Eisenhart \cite{Eisenhart} presented the metric \eqref{eq:n+1_metric} with $N \equiv 0$  `not as a special case of the more general theory, but on an independent basis' (in his own words). 
The metric \eqref{eq:n+1_metric} generalizes the one of Eisenhart to include a vector potential,  and also  clarifies its dynamical origin. 
 
Similarly to the results obtained in the previous section \ref{sec:geodesic_lift}, there is a correspondence between $\dm+1$-dimensional conserved quantities that are homogenous polynomials of order $p$ in the momenta, and $\dm$-dimensional conserved quantities that are polynomials of order $p$. In order to lift the latter to the former one follows the same procedure as in eq.\eqref{eq:conserved_quantity_lift}, making the polynomial homogeneous. $\dm+1$-dimensional homogenous conserved quantities then will correspond to Killing tensors 
 \cite{gibbons-rietdijk-vholten1993,jwvholten2006,Visinescu2009}. 
 
The  ``$\dm +2$" and  ``$\dm +1$" Eisenhart metrics
present some differences. The former has Lorentzian signature if $h_{ij}$ is Riemannian and is always well defined. The latter instead can change its signature if $\Phi$ changes sign. In particular, as  seen from eq.\eqref{eq:n+1_Hamiltonian}, the determinant of the inverse metric contains a factor $2 \Phi \det \left( h^{-1} \right)$: since $\Phi$ is defined modulo a constant, then in the neighbourhood of any point one can make $\det g_{\dm+1}^{-1} > 0$, but this can be made to work everywhere only if $\Phi$ has a global minimum. If this is not the case, then we can still use the $\dm+1$ metric to study the dynamics in a certain open subset of phase space.

\subsection{Geometrisation of the Symmetry Algebra} 

The Eisenhart-Duval lift encodes in its geometry the symmetry algebra of the lower dimensional theory. This is crucial in the non-relativistic AdS/CFT correspondence, see for example \cite{PeterChristianDuvalHassaine2008,DuvalLazzarini2012}. In this section we show that the $d$-dimensional theory admits a symmetry that maps solutions into solutions if and only if the symmetry can be lifted to a conformal transformation of the $\dm+2$-dimensional theory. In this case, a conserved charge exists in both theories and in $\dm$-dimensions takes the usual form that follows from Noether's theorem. 
 
When $h_{ij}$ is Riemannian, the Brinkmann metric \eqref{eq:Brinkmann_metric} induces a $\dm +1$ Galilean structure as follows \cite{DGH91}. $\dm + 1$-dimensional spacetime $B$ is  the quotient of the $\dm+2$-dimensional Brinkmann manifold by the action of the covariantly constant null vector $\partial_v$. The inverse metric $g^{\mu\nu}$ induces a degenerate symmetric contravariant tensor with signature $(++\dots +0)$, the projection of which 
 is given by $h^{ij}$.  $B$  also carries a closed 1-form $\theta$ called the ``absolute clock'', whose pullback to the Brinkmann manifold is $- \frac{1}{q} g (\partial_v) = - \frac{1}{q} du$. $\theta$ is in the kernel of the degenerate symmetric contravariant tensor and allows to build an absolute time variable $t$ and consider coordinates $(x^i, t)$. 
 
Let us now consider  a Galilean transformation $x \rightarrow \tilde{x}(x, t)$, $t \rightarrow \tilde{t}(t)$, or infinitesimally  
\be \label{eq:Galilei_transformation} 
\delta x^i = f^i (x, t) 
\qquad 
\delta t = f^t (t) \, ,  
\ee 
where the $t$-variation cannot depend on  $x$, and suppose that it maps solutions of the equations of motion into solutions. This means that the action 
\be 
S \left[ t_i, t_f, x(t)  \right] = \int_{t_i}^{t_f}\! L_0 \, dt = \int \left( \frac{1}{2} h_{ij} \frac{d x^i}{dt} \frac{d x^j}{dt} - q^2 \Phi - q N_i \frac{d x^i}{dt} \right) dt \,  
\ee 
can change at most by a boundary term, or, equivalently, that the integrand $L_0 \, dt$ changes by a total derivative. A detailed calculation shows that then there exists a function $g(t, x, x^\prime)$ such that 
\bea 
\frac{d g}{dt} &=&  \left. \left[ \delta L - q  \delta \dot{v} \right]  \right|_{p_v = q} \, ,  
\eea 
where $L$ is the higher dimensional Lagrangian. 
In other words, a symmetry of the $d$-dimensional equations of motion exists if and only if the transformation \eqref{eq:Galilei_transformation} can be can lifted to the whole Eisenhart-Duval space. Choosing $\delta v = {g}/{q} + \text{const.}$ the higher dimensional Lagrangian $L$ is unchanged when restricted to null geodesics. 
 
We now assume that such a lifted transformation does exist and study  its properties in some details. By asking that the Lagrangian $L$ be invariant order by order in $\dot{x}$ we find the following conditions. At second order in $\dot{x}^i$ it must be that 
\be 
\nabla^{(i} f^{k)} =  \frac{1}{2} \partial_t \left( f^t h^{ij} \right) \, , 
\ee 
which can be recognised to match the second equation in \eqref{233} upon the identifications $f^t = - \frac{1}{q} \hat{C}^u$, $f^i = \frac{1}{q} C_{(1)}^i = \hat{C}^i$. Next, to order $1$ and $0$, respectively, one finds 
\be 
\partial_i \, \delta v =  f^j \mathcal{F}_{ij} - \partial_i \left( f^j N_j \right) + \frac{1}{q} \partial_t ( f^j ) h_{ij} -  f^t \partial_t N_i \, , 
\ee  
and 
\be 
\partial_t  \, \delta v  = - q v^i \nabla_i \Phi - \partial_t \left( f^i N_i \right) + f^i \partial_t N_i + \partial_t \left( f^u \Phi \right) \, . 
\ee 
Upon using the identifications \eqref{eq:vector_identifications}, and $f^v =  \hat{C}^v  - N_i f^i$, these can be seen to match the third and, respectively fourth equations in \eqref{233}. 
 
Then, using the results of section \ref{sec:conformal_killing_vectors}, this means that the quantities $f^v$, $f^t$, $f^i$ are the components of a conformal Killing vector 
\be 
\h{K} = f^v \, \partial_v - q \, f^t \partial_u + f^i \, \partial_i 
\ee 
for the Brinkmann metric \eqref{eq:Brinkmann_metric}. The Galilean transformation \eqref{eq:Galilei_transformation} on the base $B$ lifts to a conformal transformation of the Eisenhart-Duval lift.

 For a conformal Killing vector it is possible to construct a conserved quantity $\hat{C}_{\h{K}} = \h{K}^\mu \hat{p}_\mu$. No central charges arise for tranformations generated by conformal Killing vectors  $\{ \h{K}_A \}$, since the moment maps satisfy 
\be \label{eq:no_central_charge}
\{\hat{C}_{\h{K}_A},\hat{C}_{\h{K}_B} \} = -\hat{C}_{ [\h{K}_A, \h{K}_B ] } \, . 
\ee 
The situation on the base spacetime $B$ is different and  central charges can arise: the reader may consider the Galilei group  
 which acts on $B$ but the lower dimensional moment maps realise its central extension, the Bargmann group only. 
 
Similar considerations apply to the algebraic structure of higher order conserved quantities. As it has been shown in \cite{GaryDavidClaudeHouri2011} conserved quantities of higher order in the momenta can be lifted from $\dm$-dimensional space to the Eisenhart-Duval space, and are there generated by conformal Killing tensors. 

A relationship similar to eq.\eqref{eq:no_central_charge} exists between the Poisson algebra of the moment maps and the Schouten-Nijenhuis bracket algebra of conformal Killing tensors. 
 
\section{Applications}

Now we present several systems
 with conserved quantities which are polynomial in the momenta. We derive new dynamical systems  with homogeneous conserved quantities by Eisenhart lift and obtain non-trivial Killing tensors of various ranks. 
 
\subsection{Schr\"odinger symmetry\label{sec:free_particle}} 

As an example of time-dependent conformal Killing vectors, we consider a free particle in flat three-space, described by (\ref{eq:Brinkmann_metric}) with $h^{ij} = \delta^{ij}$, $\Phi = 0$, $\mathcal{F} = 0$. Then the equations (\ref{233}) reduce to 
\bea 
\wh{C}^u = \wh{C}^u (t),
\quad \partial^{(i} C^{ \, j)}_{(1)}=-\frac{1}{2} \partial_t\wh{C}^u\delta^{ij},
\quad 
\partial^{i} C_{(0)}= - \partial_t C^{\, i}_{(1)}, \quad  \partial_t C_{(0)} = 0,  
\label{51}
\eea 
which have a solution parameterised by 13 constants $\alpha^i$, $\beta^i$, $\mu$, $\gamma_0$, $\gamma_1$, $\gamma_2$, $\delta_k$:
\be\left\{\barraynb{lll} 
\wh{C}^u &=& \gamma_0 + \gamma_1 t + \gamma_2 t^2 \, ,
\\ 
C_{(1)}^i &=& - \frac{1}{2} \left( \gamma_1 + 2 \gamma_2 t \right) x^i + \left(\alpha^i + \beta^i t \right) + \epsilon^{ijk} x_j \delta_k \, ,\\ 
C_{(0)} &=& \frac{1}{2} \gamma_2 x^2 - \beta^i x^i + \mu   \, .
\earraynb \right.
\label{52}
\ee 
$\alpha^i$ is associated with the conservation of momentum, $\beta^i$ with Galilei boosts/center of mass $t\vec{p} - \vec{x}$,  $\vec{\delta}$ with the angular momentum, $\mu$ with the mass. 
$\gamma_0$  $\gamma_1$ 
and $\gamma_2$ generate time translations, dilatations and  expansions,  with conserved charges
\beq\left\{\barraynb{llll} 
E&=&{\vec{p}^2}/{2} &\hbox{energy}
\\
D&=&t E - \frac{1}{2} \vec{x} \cdot \vec{p}
&\hbox{dilatations}
\\
K &=& t^2 E - t \vec{x} \cdot \vec{p} + \frac{1}{2} \vec{x}^2\qquad
&\hbox{expansions}
\earraynb\right.
\label{endilexp}
\eeq
respectively.
We thus recover the generators of the Schr\"{o}dinger algebra of symmetries of a free classical particle in flat spacetime, see \cite{Jackiw1972,Niederer1972,Hagen1972}. 

This result is readily extended by adding a Dirac monopole \cite{Jackiw:1980mm,Duval0}, described by
the metric
(\ref{eq:Brinkmann_metric}) with 
\beq
h^{ij} = \delta^{ij},\quad \Phi = 0, 
\quad
\mathcal{F}_{ij} = eg\frac{\epsilon_{ijk}x^k}{r^3}.
\eeq
The  $\mathcal{F}$-term  only changes the third condition in eqn (\ref{51}), which becomes 
\beq
\partial^{i} C_{(0)} = - \partial_t C^{\,i}_{(1)}
- C^{\,j}_{(1)}\mathcal{F}^{i} {}_j \, .
\eeq
The new term breaks the invariance under boosts and space translations, reducing (\ref{52}) to 
\be\left\{\barraynb{lll} 
\wh{C}^u &=& \gamma_0 + \gamma_1u + \gamma_2u^2 \, ,
\\ 
C_{(1)}^i &=&-\frac{1}{2}\left(\gamma_1 + 2\gamma_2 u \right)x^i + \epsilon^{ijk}x_j\delta_k \, ,
\\ 
C_{(0)} &=& \frac{1}{2}\gamma_2x^2 + \mu - 
eg\displaystyle\frac{\delta_i x_i}{r} \, .
\earraynb \right.
\label{54}
\ee 
Space rotations and mass-conservation-generating ``vertical" translations remain hence symmetries, but so do also time translations, dilations and expansions. The associated conserved quantities are still as in (\ref{endilexp})
up to replacing $p_i$ by the covariant expression $\Pi_i=p_i + N_i$. 

\subsection{Time dependent gravitational constant\label{sec:Gt}}

To illustrate our general theory in the case of rank 2 conformal tensors, we consider
  the ``Dirac-Vinti-Lynden-Bell" theory of gravity where the gravitational constant changes with time \cite{Dirac,Vinti-Lynden-Bell} \footnote{In an alternative theory  \cite{LBel}
  the gravitational constant is assumed to vary as
$G_0 (1 - \mathfrak{p}u_*)$.  
 For small values of the parameter $\mathfrak{p}$ the two theories yield equivalent results.}. 
 As shown in \cite{DGH91}, the
 solutions of Newton's equations with time dependent potential 
\be 
V(\vec{x}_*, u_*) = - G(u_*) \frac{M}{|\vec{x}_*|} 
\ee  

can be transformed into solutions with  time independent potential 
\be 
V(\vec{x},u) = - G_0 \frac{M}{|\vec{x}|}\, 
\ee 
by the conformal transformations 
\bea \label{eq:VLB}
\vec{x}_* = \Omega(u) \vec{x} \, , 
\quad
u_* = - \frac{a^2}{u+b} + c \, , 
\quad
v_* &=& v + \frac{\vec{x}^2}{2(u+b)} + d \, , 
\\
G(u_*) | \Omega(u_*)|= G_ 0 \,
\qquad\hbox{with}\qquad
\Omega(u)=\frac{a}{u + b}, 
\eea
where $a \neq 0$, $b$, $c$, and $d$ are constants. Notice that  $\Omega(u_*) = - {u_* - c}/{a}$.  
 Then using the transformations \eqref{eq:VLB} the standard Runge-Lenz vector of the time-independent Kepler problem,
\be 
\vec{A}_0 = ( \vec{x} \times \dot{\vec{x}} ) \times \dot{\vec{x}} + G_0 M \frac{\vec{x}}{|\vec{x}|} \, , 
\ee 
can be ``exported'' to the time-dependent theory, as 
\be \label{eq:time_dependent_RL}
\vec{A}(u_*, \vec{x}_*) =  \Omega(u_*) ( \vec{x}_* \times \vec{x}_*^{\, \prime} ) \times \vec{x}_*^{\, \prime} +  \frac{1}{a} ( \vec{x}_* \times \vec{x}_*^{\, \prime} ) \times \vec{x}_*  - \Omega(u_*)  V \vec{x} \, , 
\ee 
where $\vec{x}_*^{\, \prime} = {d \vec{x}_*}/{d u_*}$. 
Then  the quantity $\vec{A} \cdot \vec{W}$  is  conserved for any constant vector $\vec{W}$ . Comparing this with the last line of eqn \eqref{eq:rank2_ansatz} allows us to deduce that  
\bea \left\{\begin{array}{lll}
d_{(2)} &=& - 2 \Omega \, x_{*}\cdot \vec{W} \, ,  
\\[4pt] 
\wh{C}_{}^{ab} &=& 2 \Omega W^{(a} x_*^{b)} \, , %
\\[4pt] 
d_{(3)}^a &=& \displaystyle\frac{1}{a} \left( - x_*^a \, x_{*}\cdot \vec{W} + W^a x_*^2 \right) \, , 
 \\[4pt] 
d_{(4)} &=& - \Omega V q_{*} \cdot \vec{W} \, ,  
\end{array} \right. 
\eea 
and as explained in sec.\ref{sec:rank2} the conserved quantity is associated to a conformal rank $2$ Killing tensor. 
Now the equations \eqref{eq:full_rank2} reduce, for $h^{ij} = \delta^{ij}$, $\mathcal{F} = 0$, $\Phi = V$, $d_{(0)} = 0$, $d_{(1)}^i = 0$, to 
\bea  \left\{\begin{array}{cll}
\nabla^{(i} \left( d_{(2)} h^{jk)} + 
\wh{C}^{jk)} \right) &=&  0 
\label{eq:VLB_condition1}
\\ 
 \nabla^{(i} d_{(3)}^{j)} &=& - \frac{1}{2} \partial_u \left( d_{(2)} h^{ij} + 
\wh{C}^{ij} \right)   \, , 
 \label{eq:VLB_condition2} 
 \\ 
\nabla^i d_{(4)} &=& -  \partial_u d_{(3)}^{\, i}  +  \left( d_{(2)} \nabla^i \Phi + 
\wh{C}^{i l} \nabla_l \Phi \right) \, ,  
\label{eq:VLB_condition3} \\ 
0 &=& -  \partial_u d_{(4)} + q^2 d^i_{(3)} \partial_i \Phi 
\label{eq:VLB_condition4}   \, .
\end{array}\right.  
\eea 
The first equation in \eqref{eq:VLB_condition1} is satisfied as 	$d_{(2)}$ and $
\wh{C}^{ij}$ are for this problem the same as in the standard Kepler problem with an $\Omega$ rescaling. The second equation is instead a non-trivial test of our new formulae, since in the standard Kepler problem $d_{(3)}$ would be zero. It is directly seen to be satisfied once we recognize that ${d\Omega (u_*)}/{d u_*} = - {1}/{a}$. The third equation is satisfied for the same reasons it is satisfied in the standard Kepler problem, while the last condition is also satisfied once we realize that $\partial_u d_{(4)} = 0$. 
The conservation of eq. \eqref{eq:time_dependent_RL} can also be checked directly,  by calculating the time derivative.

\subsection{Lynden-Bell's Transformation of the Lorentz Force\label{sec:Lorentz}} 
 
Another similar observation, also by
Lynden-Bell \cite{LyBwandering}, concerns motion under the Newtonian force  
\begin{equation}
\vec{F}_*=q\left( \vec{v}_* \times \vec{B}_* - \frac{1}{t_*}\vec{x}_* \times \vec{B}_* \right) \,  
\label{eq:LB_generic_force}
\end{equation} 
where $\vec{B}_*(\vec{x}_*,t_*)$ is any vector.
The quantity $ \vec{x}_*-\vec{v}_* t_* $, which would be  conserved  if we had boost invariance is plainly not conserved
if the fields do not vanish.
However  its \emph{square},
\be
\left( \vec{x}_*-\vec{v}_* t_* \right) ^{2}
\label{CM2}
\ee
\emph{is conserved}.
 Such a force arises, for example, when  $\vec{B}_*$  is divergenceless,$\vnabla\cdot\vec{B}_*=0$ and the second term is the electric force, \be \label{eq:LB_electric_field}
\vec{E}_* = - \frac{1}{t_*} \vec{x}_* \times \vec{B}_* \, . 
\ee 
Faraday's equation $\vec{\nabla} \times \vec{E}_* = -  {\partial \vec{B}_*}/{\partial t_*}$ is satisfied when 
\be \label{eq:LB_magnetic_field}
\vec{B}_* = \frac{1}{(\omega t_*)^2}\vec{b} \left( \frac{\vec{x}_*}{\omega t_*} \right) \, , 
\ee 
where $\vec{b}(\vec{x})$ is any divergenceless vector, and $\omega$ is a constant with the dimension of inverse time. 
Lynden-Bell then showed that the following change of coordinates 
\be 
\label{eq:LB_space_time_transformation_specific}
  \vec{x} = \displaystyle\frac{\vec{x}_*}{\omega t_*} \, 
\qquad
  t = \displaystyle\frac{1}{\omega^2 t_*}  
\ee  
transforms motion under the force \eqref{eq:LB_generic_force}  
into one 
 and 
\be 
\label{eq:LB_magnetic_transformation_specific}
\vec{F} = q\, \vec{v}\times \vec{B},
\qquad
\vec{B}( \vec{x}, t) = - (\omega t_*)^2 \vec{B}_* (\vec{x}_*, t_*) \, . 
\ee 
where $\vec{v}={d\vec{x}}/{d t}$.
In particular when $\vec{B}_*$ is given by equation \eqref{eq:LB_magnetic_field} then $\vec{B}(\vec{x},t) = - \vec{b}(\vec{x})$, a generic stationary magnetic field. 
 
Lynden-Bell then goes on to say that there is a strange transformation that can be applied to any pair of electric and magnetic fields $\vec{E}_*$, $\vec{B}_*$ that satisfy the sourceless Maxwell equations. The transformation is given by \eqref{eq:LB_space_time_transformation_specific}, \eqref{eq:LB_magnetic_transformation_specific} supplemented with 
\be \label{eq:LB_electric_transformation_specific}
\vec{E}(\vec{x}, t) = (\omega t_*)^3 \left[ \vec{E}_* + \frac{1}{t_*} \vec{x}_* \times \vec{B}_* \right] \, . 
\ee 
Then the equations of motions  are unchanged, 
\be 
\frac{d^2\vec{x}}{dt^2} = q \left(  \vec{v} \times \vec{B} + \vec{E} \right) \quad \Leftrightarrow \quad \frac{d^2\vec{x}_*}{dt_*^2} = q \left(  \vec{v}_* \times \vec{B}_* + \vec{E}_* \right) \, . 
\ee 
 
In this section we show how Lynden-Bell's observations can be generalised and interpret them in terms of conformal mappings of the Brinkmann metric. Then, for the cases where \eqref{eq:LB_magnetic_field} holds the quantity in (\ref{CM2}) 
 provides a new example of an explicitly time dependent conserved quantity for the Brinkmann metric, associated with an explicitly time-dependent Killing tensor of rank two.
 
We begin with the Lorentz force equation 
\be \label{eq:Lorentz_force} 
m \frac{d^2 \vec{x}_*}{dt_* ^2 } = q \Bigl(\vec{E}_* + \frac{d \vec{x}_*}{dt_*} \times \vec{B}_*  \Bigr ) 
\ee
and consider the general transformation 
\be
t_*=f( t) \,,\qquad \vec{x}_*= \sqrt{|f^\prime|} \vec{x} \, ,  \label{gentrans}
\ee
where $f^\prime = {df}/{dt} \neq 0$. 
 The velocity transforms as 
\be
\frac{d \vec{x}_*}{dt_*} = \frac{\sgn(f^\prime)}{\sqrt{|f^\prime|}} 
\frac{d \vec{x}}{d t} + \frac{1}{2} \frac{f^{\prime \prime}} {|f^\prime|^{\frac{3}{2}}} 
\vec{x},
 \label{velocity} 
\ee
so  one gets 
\be \label{eq:LB_electric_transformation}
m \frac{d^2 \vec{x}}{d t ^2 }
+ \frac{1}{2} \sgn (f^\prime) \sqrt{|f^\prime|}  \left( \frac{f^{\prime \prime}}{|f^\prime|^\frac{3}{2} } \right)^\prime m  \vec{x} = q ( \vec{E} + \frac{d \vec{x}}{d t} \times \vec{B}), 
\ee
 where 
\be \label{eq:LB_magnetic_transformation}
\vec{E} = |f^\prime|^{\frac{3}{2}} \vec{E}_* + 
\frac{1}{2}   \frac{f^{\prime \prime}}{\sqrt{|f^\prime|}}  \vec{x}_* \times \vec{B}_* \,,\qquad \vec{B} = f^\prime  \vec{B}_* \,. 
\ee 
In particular, for $f(t) = {1}/{\omega^2 t}$ one recovers \eqref{eq:LB_magnetic_transformation_specific}, \eqref{eq:LB_electric_transformation_specific}. 
 Notice that 
\be \label{eq:Schwarzian_pre} 
\frac{1}{2} \sgn (f^\prime) \sqrt{|f^\prime|}  \left( \frac{f^{\prime \prime}}{|f^\prime|^\frac{3}{2} } \right)^\prime  = \frac{1}{2} \{ f, t \} \, , 
\ee
where
\be \label{eq:Schwarzian}
\{ f, t \} = \Bigl( \frac{f^{\prime \prime}} {f^\prime } \Bigr )
 ^\prime - \frac{1}{2}  \Bigl( \frac{f^{\prime \prime}}  {f^\prime}  \Bigr ) ^2 = 
\frac{f^{\prime \prime \prime} } {f^\prime} - \frac{3}{2}
\Bigl( \frac{f^{\prime \prime} } {f^\prime }  \Bigr ) ^2  
\ee 
is the Schwarzian derivative. The latter is known to vanish if and only if
$f$ is a fractional linear transformation
\be
\{f, t \} =0 \,, \qquad \Longleftrightarrow   \qquad f= \frac{A t + B}{C  t  + D } \,, \quad AD-BC \ne 0 \, . 
\label{frac}\ee 
Then when $\{f, t \} = 0$ (which includes Lynden-Bell's observation), then no linear term in $\vec{x}$ is induced
(\ref{eq:LB_electric_transformation}).

Now suppose $\{f, t \} = 0$. Then if  $\vec{E} = 0$, then the square of the quantity in  (\ref{velocity}) is a constant. Using eq. \eqref{eq:Schwarzian_pre} and assuming $f^{\prime\prime} \neq 0$ we infer that 
\be 
\left( \vec{x}_* - 2 \frac{|f^\prime|^2}{f^{\prime\prime}} \frac{d \vec{x}_*}{d t_*}\right)^2 
\ee 
is constant. In Lynden-Bell's example this is precisely $(\vec{x}_* - \vec{v}_* t_*)^2$ in (\ref{CM2}). As discussed in sec.\ref{sec:conformal_killing_vectors} this conserved quantity is associated to a conformal Killing vector.  
 
Eqns \eqref{eq:LB_electric_transformation}, \eqref{eq:LB_magnetic_transformation} imply that for  $F = - E_i dt \wedge dx^i + \frac{1}{2} \epsilon_{ijk} dx^i dx^j B_k$ one has 
$ 
F_* = \sgn(f^\prime) F \, . 
$ 
If
 we  introduce the potential $A = q \, \Phi \, d t + N_i d x^i$, then 
\be  
q \, \Phi = q \, |f^\prime| \Phi_* + 
\displaystyle\frac{f^{\prime\prime}}{2 |f^\prime|} x_*^i N_{* i} \, , 
\qquad 
N_i = \sgn(f^\prime) \sqrt{|f^\prime|} N_{* i} \, , 
\label{qPhiN}  
\ee 
which yields the  transformation of the electric and magnetic fields. 
Then using \eqref{qPhiN} we see that (\ref{gentrans}) induces a conformal transformation of the Bargmann metric. Writing the latter  as  
\bea 
ds_*^2 = g_{*\mu\nu} dx^\mu dx^\nu &=& h_{ij} dx_*^i dx_*^j + 
 2 du_* \left( dv_* - \Phi_* du_* + N_{* i} dx_*^i \right) \nonumber \\ 
&=& h_{ij} dx_*^i dx_*^j -  
 2 q dt_* \left( dv_* + A_* \right)  \, . 
\eea
where $u = - q dt$, and using  \eqref{gentrans} gives
\bea 
\hspace{-0.75cm} ds_*^2 = |f^\prime| \left[  h_{ij} d x^i d x^j - 
 2 q d t \left( \sgn (f^\prime) dv_* + A \right) + \left( \frac{f^{\prime\prime}}{2f^\prime}\right)^2 h_{ij} x^i x^j d t^2 + \frac{f^{\prime\prime}}{f^\prime} h_{ij}  x^i d x^j d t \right].  
\eea 
Defining $v_* = \sgn( f^\prime ) \left( v + \frac{1}{4q} \frac{f^{\prime \prime}}{f^\prime} h_{ij} x^i x^j \right)$ this becomes 

\be 
ds_*^2 = |f^\prime| \left[  h_{ij} dx^i d x^j - 
 2 q d t \left( d v + A \right) \right] = |f^\prime| ds^2 \,,  
\ee 
as stated. Because two conformally related metrics have the same null geodesics up to a change of parameterisation, this explains the correspondence between solutions of the equations of motion found above.

\subsection{The Quantum Dot}\label{QDsec} 
In the case of Quantum Dots the relative motion of two electrons can be described by the Bargmann metric
\beqa
d\vY^2+2du(dv-\half B\rho^2 d\varphi)-2\Phi(\vY)du^2,\quad
\Phi(\vY)=\frac{1}{2}\Big(\omega^2_0\rho^2+\omega_z^2z^2\Big)-\frac{a}{\sqrt{\rho^2+z^2}},
\qquad
\label{decompmet}
\eeqa
where  $\vY= (\rho, \varphi,z)$ is the relative position in cylindrical coordinates and $B$ a constant magnetic field. (\ref{decompmet}) corresponds to an anisotropic axially symmetric oscillator combined with 
a Coulomb potential and a uniform magnetic field.    
Non-trivial conserved quantities arise when the parameter
\be
\tau = \frac{\omega_z}{\sqrt{\omega_0^2 + \frac{B^2}{4}}},
\label{tau}
\ee  
 which measures the extent of anisotropy, takes special values
 $\tau=1,2, 1/2$. For $\tau=1$ we have rotational symmetry and the conserved angular momentum corresponds to a Killing vector; for $\tau=2$  there is 
a quadratic conserved quantity $Q_1$ associated to a rank two Killing tensor, which generalises the $z$-component of the  Runge-Lenz vector of the pure Coulomb problem  \cite{ZZHG,CGvHHKZ2014}. Lifted to the Bargmann space, it reads
\be 
\h{Q}_1 = z \Pi_{\rg}^2 - \rg \Pi_{\rg} \Pi_z + \frac{z}{\rg^2} \Pi_{\vf}^2 + B z \Pi_{\vf} p_v- \left(\omega_0^2 \rg^2 z + \frac{az}{\sqrt{\rg^2 +z^2}} \right) p_v^2. 
\label{Q1} 
\ee 
For  $\tau = \frac{1}{2}$ instead a quartic conserved quantity associated to a rank four Killing tensor arises: this is $Q_2$  in 
eqn. \# (3.24) of Ref. \cite{CGvHHKZ2014} (not reproduced here). In this case 
 the system is integrable but not separable. For  details,
 reader is  referred to 
\cite{Zhang11, ZZHG}.

\subsection{The H\'enon-Heiles system}\label{HHsec}

In 1964 H\'enon and Heiles (HH) \cite{HenonHeiles1964} proposed 
to describe stellar dynamics in an axially symmetric galaxy by the Hamiltonian with a  cubic potential, 
\beq
H= \frac12 \left(p_1^2 + p_2^2 +\omega_1 q_1^2 + \omega_2 q_2^2\right) + 
\alpha\, q_1 q_2^2  - \frac{1}{3}\beta\, q_1^3, 
\label{H2Ham}
\eeq
where $q_1,q_2$ are coordinates in the galactic plane and $p_1$ and $p_2$ are the associated momenta.

The original HH system has
$
\omega_1 =\omega_2 =\alpha=\beta=1\,
$ 
and a chaotic behavior.   
The Hamiltonian is Liouville integrable in three cases \cite{BountisSegurVivaldi1982,ChangTaborWeiss1982,GrammaticosDorizziPadjen1982} \footnote{These cases arise 
by reduction from the Sawada-Kotera, $KdV_5$ and Kaup-Kupershmidt systems, respectively \cite{Fordy1991}, as hinted at by our notations. What is now known as the generalised H\'enon-Heiles system is obtained by adding $\frac{\gamma}{2 q_2^2}$ to the Hamiltonian  (\ref{H2Ham}) \cite{VerhoevenMusetteConte2005}. In this paper we study the case $\gamma=0$ only.},  
namely for
\beq
\left\{\begin{array}{ccll}
(i)\qquad  &(SK)\qquad &\beta/\alpha =-1,\, \qquad &\omega_1=\omega_2\, 
\\[4pt]
(ii)\qquad &(KDV5)\qquad&\beta/\alpha =-6\,
& 
\\[4pt] 
(iii)\qquad  &(KK)\qquad&\beta/\alpha= -16\,, \quad\quad &\omega_1=16\omega_2
\end{array}\right.
\label{goodvalues}
\eeq

The Eisenhart lift of the additional constants of the motion is given by 
\bea 
(SK)\quad  \; K^{(i)} &=&\left[3p_1p_2 + \alpha q_2(3q_1^2 + q_2^2) \h{p}_v^2 + 3\omega_2q_1 q_2 \h{p}_v^2\right]^2 
\\[6pt] 
(KDV5)\quad \; K^{(ii)} &=&4\alpha p_2\left(q_2 p_1 - q_1 p_2 \right) + (4\omega_2-\omega_1)(p_2^2+\omega_2q_2^2 \h{p}_v^2)  \nn 
\\ && 
+ \alpha^2q_2^2(4 q_1^2 + q_2^2 )\h{p}_v^2 + 4\alpha q_1 \omega_2q_2^2 \h{p}_v^2  
\\[6pt] 
(KK)\quad \; K^{(iii)} &=& \left[3p_2^2 + (3\omega_2 q_2^2 
) \h{p}_v^2 \right]^2 + 12\alpha p_2 q_2^2 ( 3 q_1 p_2 - q_2 p_1 ) \h{p}_v^2 \nn 
\\ 
&& + \left[-2\alpha^2 q_2^4 (6 q_1^2 + q_2^2 ) - 12\alpha q_1 \omega_2 q_2^4\right] \h{p}_v^4  \, .
\eea 
The first two correspond to rank $2$ Killing tensors, and the third to a rank $4$ Killing tensor. 
\subsection{The Holt system}
 
The Holt system \cite{Holt1982} 
has two degrees of freedom, described by the Hamiltonian  
\be 
H = \frac{1}{2} \left( p_x^2 + p_y^2 \right) + \frac{3}{4} \mu x^{\frac{4}{3}} + y^2 x^{ - \frac{2}{3}} \, . 
\ee 
The system is integrable for $\mu = 1, 6, 16$ \cite{GrammaticosDorizziRamani1984,GrammaticosDorizziRamaniHietarinta1985}; the associated conserved quantities are of order $3$, $4$ and, respectively $6$. Setting $q=1$, the Eisenhart lifts are, accordingly: 
\bea 
(\mu = 1) \quad 
\wh{C} &=& p_y^3 + \frac{3}{2} p_y \, p_x^2 + \left(- \frac{9}{2} x^{\frac{4}{3}} + 3 x^{- \frac{2}{3}} y^2 \right) p_y \, {\h{p}_v}^2 + 9 \, x^{\frac{1}{3}} \,  y\,  p_x\, {\h{p}_v}^2 \, , \nn 
\\[6pt] 
(\mu = 6) \quad 
\wh{C} &=& p_y^4 + 2 \, p_y^2 \, p_x^2 + 4 \, x^{- \frac{2}{3}} y^2 p_y^2 \, {\h{p}_v}^2 + 24 \, x^{\frac{1}{3}} y \, p_y \, p_x \, {\h{p}_v}^2 + 72 \, x^{\frac{2}{3}} y^2 {\h{p}_v}^4 \, , \nn 
\\[8pt] 
(\mu = 16) \quad 
\wh{C} &=& p_y^6 + 3 p_y^4 \, p_x^2 + \left(18 x^{\frac{4}{3}} + 6 x^{- \frac{2}{3}} y^2 \right) p_y^4 \, {\h{p}_v})^2 + 72 \, x^{\frac{1}{3}} y \, p_y^3 \, p_x \, {\h{p}_v}^2 \nn 
\\[8pt]  
&& + 648 \, x^{\frac{2}{3}} y^2 p_y^2 \, {\h{p}_v}^4  + 648 y^4 {\h{p}_v}^6 \, , 
\eea  
providing us with higher-order Killing tensors of rank $3$, $4$ and, respectively, $6$. 

In their three  integrable cases the Holt system and  H\'enon-Heiles systems are related by a (non-canonical) duality transformation
  \cite{Hietarinta1983,HietarintaGrammaticosDorizziRamani1984}.

\section{Conclusion}

In this paper we have given a covariant algorithm for deriving conserved quantities for natural Hamiltonian systems and shown how they may be constructed from conformal Killing tensors in the associated higher dimensional (Lorentzian or Riemannian) manifold(s) in which the dynamical trajectories lift to (null) geodesics. We have described in detail the role of explicit time dependence and shown how this is related to lifted Killing tensors that are properly conformal. We have illustrated the general theory by several examples, a number of them explicitly time-dependent.  

\begin{acknowledgments}
The authors are indebted to Piotr Ko\'sinski for his interest and advice.
GWG is grateful to the 
{\it Laboratoire de Math\'ematiques et de Physique Th\'eorique de l'Universit\'e de Tours}  for hospitality, and the  R\'egion Centre for a 
\emph{``Le Studium''} research professor\-ship.
For JWvH this work is part of the research program of the Foundation for Research of Matter (FOM).
PH and PMZ would like to thank the {\it Institute of Modern Physics}
at the Lanzhou branch of the Chinese Academy of Sciences and the {\it Laboratoire de Math\'ematiques et de Physique Th\'eorique de l'Universit\'e de Tours}, respectively, for hospitality.
 This work was partially supported also by the National Natural Science Foundation of
China (Grants No. 11035006 and 11175215) and by the Chinese Academy of Sciences Visiting
Professorship for Senior International Scientists (Grant No. 2010TIJ06). 
\end{acknowledgments} 
\goodbreak


\vskip5mm
\appendix{\bf Appendix:  Christoffel symbols of 
the d+2  metric}


\bigskip

The non-zero Christoffel symbols for the $\dm+2$ lift metric \eqref{eq:Brinkmann_metric} are 

\be 
\left\{ 
\begin{array}{lcl}  
\h{\Gamma}^v_{uu} &=& - N_i \, \partial^i \Phi  -  \partial_u \left( 2 \Phi - \frac{1}{2} N^2 \right)   \, , \nn \\ 
\h{\Gamma}^v_{i u} &=& \frac{1}{2}  N_l F^l_{\;\; i} - \partial_i \Phi + \frac{1}{2} \partial_u ( N^l) h_{li} \, , \nn \\  
\h{\Gamma}^v_{i j} &=& \nabla_{(i} N_{j)}  \, , \nn \\  
\h{\Gamma}^i_{uu} &=& \partial^i \Phi  + h^{il} \partial_u N_l  \, , \nn \\ 
\h{\Gamma}^i_{j u} &=& -\frac{1}{2} F^i_{\;\;j} + \frac{1}{2} h^{il} \partial_u h_{lj}   \, , \nn \\ 
\h{\Gamma}^i_{j k} &=& \Gamma^i_{j k} \, . 
\end{array} 
\right. 
\ee


\begin{thebibliography}{99}
\bibitem{Benenti2001} 
S. Benenti, C. Chanu and G. Rastelli, ``Variable separation for natural Hamiltonians with scalar and vector potentials on Riemannian manifolds", J. Math. Phys. \textbf{42}(5) 2065-2091 (2001). 
  
\bibitem{KalninsMiller1980} 
E. G. Kalnins, W. Miller, 
``Killing tensors and variable separation for Hamilton-Jacobi and Helmholtz equations", 
{\it SIAM J. Math. Anal.}, 11 (1980) 1011-1026. 
 
\bibitem{KalninsMiller1981} 
E. G. Kalnins, W. Miller, 
``Killing Tensors and Nonorthogonal Variable  Separation for Hamilton Jacobi Equations", 
\textit{ SIAM J. Math. Anal.}, 12 (1981) 617-629. 
 
\bibitem{Benenti1997} 
S. Benenti, 
``Intrinsic characterization of the variable separation in the Hamilton Jacobi equation," 
{\it J. Math. Phys.} 38 (1997) 6578-6602.  

\bibitem{Benenti2005} 
S. Benenti, C. Chanu, G. Rastelli, 
``Variable-separation theory for the null Hamilton--Jacobi equation," 
{\it J. Math. Phys.}, 46 (2005) 042901.
  
\bibitem{Carter}
  B.~Carter,
 ``Global structure of the Kerr family of gravitational fields,''
  Phys.\ Rev.\  {\bf 174} (1968) 1559.

\bibitem{Crampin} 
M. Crampin,  ``Hidden symmetries and Killing tensors.",  Rep. Math. Phys. 20.1 (1984): 31-40. 
 
\bibitem{DuvalValent}
  C. Duval and G. Valent,
 ``Quantum integrability of quadratic Killing tensors,''
  J.\ Math.\ Phys.\  {\bf 46} (2005) 053516.
[arXiv:math-ph/0412059] 

\bibitem{GaryDavidClaudeHouri2011}  
G. W. Gibbons, T. Houri, D. Kubiz\v{n}\'ak, and C. M. Warnick, ``Some spacetimes with higher rank Killing-St\"{a}ckel tensors", Physics Letters B 700(1) 68-74 (2011) [arXiv:1103.5366 [gr-qc]]. 

\bibitem{Rugina}
  G.~W.~Gibbons and C.~Rugina,
 ``Goryachev-Chaplygin, Kovalevskaya, and Brdi\v{c}ka-Eardley-Nappi-Witten pp-waves spacetimes with higher rank St\"{a}ckel-Killing tensors,''
  J.\ Math.\ Phys.\  {\bf 52} (2011) 122901,  [arXiv:1107.5987 [gr-qc]].
  
\bibitem{Galajinsky}
  A.~Galajinsky,
``Higher rank Killing tensors and Calogero model,''  
Phys.\ Rev.\ D {\bf 85} (2012) 085002, [arXiv:1201.3085 [hep-th]].
   
\bibitem{Visinescu}
  M.~Visinescu,
  ``Higher order first integrals, Killing tensors and Killing-Maxwell system,''
  J.\ Phys.\ Conf.\ Ser.\  {\bf 343} (2012) 012126.

\bibitem{CGvHHKZ2014} 
  M.~Cariglia, G.~W.~Gibbons, J.-W.~van Holten, P.~A.~Horvathy, P.~Kosinski and P.-M.~Zhang,
  ``Killing tensors and canonical geometry,''
  Class.\ Quant.\ Grav.\  {\bf 31} (2014) 125001
  [arXiv:1401.8195 [hep-th]].
 
\bibitem{Marco2012} 
M. Cariglia, ``Hidden symmetries of Eisenhart-Duval lift metrics and the Dirac equation with flux", Physical Review D \textbf{86} 084050 (2012). 
   
\bibitem{gibbons-rietdijk-vholten1993} G.\ Gibbons, R.\ Rietdijk and J.W.\ van Holten, ``SUSY in the sky,'' 
Nucl.\ Phys.\ B404 (1993), 42; arXiv:hep-th/9303112v1 
 
\bibitem{jwvholten2006} J.W.\ van Holten, ``Covariant hamiltonian dynamics,''  
Phys.\ Rev.\ D75 025027 (2007), [arXiv:hep-th/0612216v2]

\bibitem{PeterNgome2009}  
P. A. Horv\'athy and J-P. Ngome, ``Conserved quantities in non-abelian monopole fields", Phys. Rev. D {\bf 79} 12 (2009) 127701. 
[arXiv:0902.0273 [hep-th]]
 
\bibitem{Ngome}
  J.~-P.~Ngome,
 ``Curved manifolds with conserved Runge-Lenz vectors,''
  J.\ Math.\ Phys.\  {\bf 50} (2009) 122901
  [arXiv:0908.1204 [math-ph]].
  
\bibitem{Visinescu2009} 
M. Visinescu, ``Higher order first integrals of motion in a gauge covariant Hamiltonian framework", Mod. Phys. Lett. A 25.05 341-350 (2010). 
  ``Covariant approach of the dynamics of particles in external gauge fields, Killing tensors and quantum gravitational anomalies,''
  SIGMA {\bf 7} (2011) 037
  [arXiv:1102.0095 [hep-th]].
   
\bibitem{Igata}
Takahisa Igata, Tatsuhiko Koike, Hideki Ishihara,
``Constants of Motion for Constrained Hamiltonian Systems,''
Phys. Rev. \textbf{D 83} 065027, (2011).
[arXiv:1005.1815]
 
\bibitem{Kubiznak}
  D.~Kubiznak and M.~Cariglia,
 ``On Integrability of spinning particle motion in higher-dimensional black hole spacetimes,''
  Phys.\ Rev.\ Lett.\  {\bf 108} (2012) 051104
  [arXiv:1110.0495 [hep-th]].

\bibitem{PeterJWNgome2010}  
J.-P. Ngome, P. A. Horv\'athy and J.-W. Van Holten, ``Dynamical supersymmetry of the spin particle--magnetic field interaction", J. Phys. A \textbf{43} 28 (2010) 285401  [arXiv:1003.0137 [hep-th]].

\bibitem{Eisenhart} 
L. P. Eisenhart, ``Dynamical trajectories and geodesics", Annals. Math. {\bf 30} 591-606 (1928). 
 
\bibitem{Dothan} 
Y. Dothan, ``Finite-Dimensional Spectrum-Generating Algebras", Phys. Rev. D {\bf 12} 2944-2954 (1970). 


\bibitem{Brinkmann1925} 
H. W. Brinkmann, 
 ``Einstein spaces which are mapped conformally on each other", Math. Ann. 94.1 119-145 (1925)
 
\bibitem{DBKP} 
C. Duval, G. Burdet, H. P. K\"{u}nzle and M. Perrin, ``Bargmann structures and Newton-Cartan theory", Phys. Rev. D {\bf 31} (1985) 1841. 

\bibitem{DGH91} 
C. Duval, G. W. Gibbons, P. Horv\'athy, ``Celestial mechanics, conformal structures and gravitational waves", Phys. Rev. D {\bf 43} 3907 (1991), [\href{http://xxx.lanl.gov/abs/hep-th/0512188}{\texttt{ hep-th/0512188}}].

\bibitem{SSD}
J.-M.~Souriau,
\textsl{Structure des syst\`emes dynamiques}, Dunod (1970, \copyright 1969);
\textsl{Structure of Dynamical Systems. A Symplectic View of Physics},
Birkh\"auser (1997).
 
\bibitem{DHP2}  
C. Duval, P. A. Horv\'athy and L. Palla, 
``Conformal properties of Chern-Simons vortices in external fields'',  Phys. Rev. {\bf D50}  (1994) 6658.  
 
\bibitem{Jackiw1972} 
R. Jackiw, ``Introducing scale symmetry", Phys. Today \textbf{25} 1 23-27 (1972). 
 
\bibitem{Niederer1972} 
U. Niederer, ``The maximal kinematical invariance group of the free Schr\"{o}dinger equations with arbitrary potentials", Helv. Phys. Acta \textbf{45} 802 (1972) 
 
\bibitem{Hagen1972} 
C. R. Hagen, ``Scale and conformal transformations in Galilean-covariant field theory", Phys. Rev. \textbf{5} 2 377 (1972).  

\bibitem{Duval0}
C.~Duval,
``Quelques proc\'{e}dures g\'eom\'etriques en dynamique des particules,''
Th\`ese d'Etat  (Marseille 1982 - unpublished).
  
\bibitem{Dirac}
P. A. M. Dirac,
``A new basis for cosmology,''
Proc. Roy. Soc. A {\bf 165} (1938) 199-208

\bibitem{DHP1}
C. Duval, P. A. Horv\'athy and L. Palla, 
``Conformal symmetry of the coupled Chern-Simons and gauged non-linear 
 Schr\"odinger equations,''  
Phys. Lett. {\bf B325}, 39 (1994) [hep-th/9401065];
 ``Spinors in non-relativistic Chern-Simons electromagnetism,''   
Ann. Phys. (N. Y.) {\bf 249}, 265 (1996) [hep-th/9510114].

\bibitem{HHhydro}
M. Hassa\"\i ne and P. A. Horv\'athy,
``Field--dependent symmetries of a non-relativistic fluid model,''
Ann. Phys. (N. Y.)  {\bf 282}, 218 (2000). 

\bibitem{PeterChristianDuvalHassaine2008} 
C. Duval, M. Hassa\"{i}ne P. A. Horv\'athy, ``The geometry of Schr{\"o}dinger symmetry in non-relativistic CFT", Ann. Phys. \textbf{324}(5) 1158--1167 (2009). 
  
\bibitem{DuvalLazzarini2012} 
C. Duval, S. Lazzarini, ``Schr\"odinger manifolds", 
J.\ Phys.\ A {\bf 45} (2012) 395203
  [arXiv:1201.0683 [math-ph]].
 
\bibitem{GaryMarco2013} 
M. Cariglia, and G. Gibbons, ``Generalised Eisenhart lift of the Toda chain", J. Math. Phys. \textbf{55} 022701 (2014). 

\bibitem{Vinti-Lynden-Bell}
J. P. Vinti,
``Classical solution of the two-body problem if the gravitational constant
diminishes inversely with the age of the universe,'' 
Mon. Not. R. Astr. Soc. {\bf 169} (1974) 417-427;
D. Lynden-Bell,
``On the $N$-Body Problem in Dirac's Cosmology,'' 
Notes from Observatories {\bf 102} (1982) 86-87

\bibitem{LBel}
L.~Bel,
 ``Earth and Moon orbital anomalies,''
  arXiv:1402.0788 [gr-qc].
  
\bibitem{LyBwandering}
D. Lynden-Bell,
 ``The Newton wonder in mechanics", Observatory {\bf 120}
(2000) 131: 
``Wandering among Newton wonders'',
Observatory {\bf 120} (2000) 192

\bibitem{DavidPavelValeriMarco2012} 
M. Cariglia, V. P. Frolov, P. Krtou\v{s} and D. Kubiz\v{n}\'ak, ``Geometry of Lax pairs: Particle motion and Killing-Yano tensors", Phys. Rev. {\bf D}  87(2)  024002 (2013).  

\bibitem{Jackiw:1980mm}
  R.~Jackiw,
``Dynamical Symmetry of the Magnetic Monopole,''
  Annals Phys.\  {\bf 129} (1980) 183.

\bibitem{QDots}
N. S. Simonovi\'{c} and R. G. Nazmitdinov,
``Hidden symmetries of two-electron quantum dots in a magnetic field,''
Phys. Rev. {\bf B67}, 041305(R) (2003); 
Y. Alhassid, E. A. Hinds and D. Meschede,
``Dynamical Symmetries of the Perturbed Hydrogen Atom~: the van der Waals Interaction,''
Phys. Rev. Lett. {\bf 59} (1987) 1545;
K. Ganesan and M. Lakshmanan,
``Comment on ``Dynamical Symmetries of the Perturbed Hydrogen Atom~: the van der Waals Interaction'''',
Phys. Rev. Lett. {\bf 629} (1989) 232.
%
R. Bl\"umel, C. Kappler, W. Quint, and J. Walter,
``Chaos and order of laser-cooled ions in a Paul trap,'' {\it Phys. Rev.} {\bf A40}, 808 (1989);
Erratum: Phys. Rev. {\bf A 46}, 8034 (1992). 

\bibitem{trap1}
 P.A. Maksym and T. Chakraborty, 
 ``Quantum Dots in a Magnetic
Field: Role of Electron-Electron Interactions,''
 Phys. Rev. Lett. {\bf 65} (1990) 108.

\bibitem{trap2} 
J.L. Birman, R.G. Nazmitdinov, V.I. Yukalov, 
``Effects of
symmetry breaking in finite quantum systems,''
 Phys. Rep. 526 (2013) 1-91. [arXiv:1305.5131].

\bibitem{ZZHG} 
P.-M.\ Zhang, L.-P.\ Zou, P.A.\ Horvathy and G.W.\ Gibbons, 
``Separability and Dynamical Symmetry of Quantum Dots,''   
  Annals of Physics (N.Y.) {\bf 341}, 94 - 116 (2014) [arXiv:1308.3035 [hep-th]].

\bibitem{Zhang11}
 P.~M.~Zhang and P.~A.~Horvathy
``Kohn's theorem and Galilean symmetry,'' 
Phys. Lett. {\bf B702} (2011) 177 
 [arXiv:1105.4401 [hep-th]].

\bibitem{HenonHeiles1964} M. H\'enon and C. Heiles, ``The Applicability of the Third Integral of Motion: Some Numerical Experiments'', Astr. J. \textbf{69}(1) 73-79 (1964). 
 
\bibitem{VerhoevenMusetteConte2005} R. Conte, M. Musette and C. Verhoeven, ``Explicit integration of the H\'enon-Heiles Hamiltonians'', J. Nonl. Math. Phys. \textbf{12}(1) 212-227 (2005). 
 
\bibitem{BountisSegurVivaldi1982} T. Bountis, H. Segur, and F. Vivaldi, ``Integrable Hamiltonian systems and the Painlev{\'e} property'', Phys. Rev. A \textbf{25}(3) 1257 (1982). 
 
\bibitem{ChangTaborWeiss1982} 
Y. Chang and M. Tabor and J. Weiss, 
``Analytic Structure of the H{\'e}non--Heiles Hamiltonian in integrable and nonintegrable regimes'', 
J. Math. Phys. \textbf{23} 531 (1982). 
 
\bibitem{GrammaticosDorizziPadjen1982} B. Grammaticos, B. Dorizzi and R. Padjen, ``Painlev{\'e} property and integrals of motion for the H{\'e}non-Heiles system'', Phys, Lett. A \textbf{89}(3) 111-113 (1982). 
  
\bibitem{Fordy1991} A. P. Fordy, ``The H{\'e}non-Heiles system revisited'', Phys. D \textbf{52}(2) 204-210 (1991). 
 
\bibitem{Holt1982} C. R. Holt, ``Construction of new integrable Hamiltonians in two degrees of freedom'', J. Math. Phys. \textbf{23}(6) 1037-1046 (1982). 
 
\bibitem{GrammaticosDorizziRamani1984} 
B. Grammaticos, B. Dorizzi, and A. Ramani, ``Hamiltonians with high order integrals and the "weak Painlev\'e" concept'', J. Math. Phys. \textbf{25} 3470-3473 (1984). 
 
\bibitem{GrammaticosDorizziRamaniHietarinta1985} 
B. Grammaticos, B. Dorizzi, A. Ramani, and J. Hietarinta, ``Extending integrable hamiltonian systems from 2 to N dimensions'', Phys. Lett. A \textbf{109}(3) 81-84 (1985). 
 
\bibitem{Hietarinta1983} J. Hietarinta, ``Integrable families of H\'enon-Heiles-type Hamiltonians and a new duality'', Phys. Rev. A \textbf{28} 3670-3672 (1983). 
 
\bibitem{HietarintaGrammaticosDorizziRamani1984} 
J. Hietarinta, B. Grammaticos, B. Dorizzi and A. Ramani, ``Coupling-Constant Metamorphosis and Duality between Integrable Hamiltonian Systems'', Phys. Rev. Lett. \textbf{53}  1707-1710 (1984). 
 
\end{thebibliography}
\end{document}